\documentclass[reprint,superscriptaddress,showpacs,preprintnumbers,notitlepage,amsmath,amssymb,onecolumn,prb,]{revtex4-1}

\usepackage{comment}
\usepackage{graphicx}
\usepackage{adjustbox}

\usepackage[usenames,dvipsnames]{xcolor}
\usepackage[colorlinks, breaklinks=true, linkcolor=red, citecolor=blue, linktocpage=true]{hyperref}
\usepackage{mathrsfs}
\usepackage{pst-node}
\usepackage{tikz-cd} 
\usepackage{cancel}
\usepackage{extarrows}
\usepackage{slashed}
\usepackage{amsmath}
\usepackage{subcaption} 
\usepackage{geometry} 

\begin{document}

\title{Constraining Ultra-Light Dark Matter mass with Dwarf Galaxy Rotation Curves}

\author{Tian-Yao Fang}
\email{tianyaofang@cuhk.edu.hk}
\affiliation{Chinese University of Hong Kong}
\author{Ming-Chung Chu}
\email{mcchu@phy.cuhk.edu.hk}
\affiliation{Chinese University of Hong Kong}

\date{\today}

\begin{abstract}
 While ultra-light bosonic dark matter (ULDM)  in a Bose-Einstein condensate (BEC) state could naturally account for  the central core  in some galaxies and resolve the core-cusp problem, the dark matter density distribution in the outer regions of galaxies remains less explored. We propose a trial wavefunction to model the ULDM distribution beyond the BEC core. We derive the corresponding rotation velocity curve, which shows excellent agreement with those of  12 dwarf spheroidal galaxies.  The best-fit ULDM particle mass for each dwarf galaxy falls  within a strikingly narrow range of $m=(1.8-3.2)\times 10^{-23}\text{eV}$.

\end{abstract}

\maketitle

\section{Introduction}

The Lambda-Cold Dark Matter  ($\Lambda$CDM) model successfully describes cosmic structure formation on large scales. However, it faces persistent challenges on galactic scales ($\le 10\ \text{kpc}$), most notably the core-cusp problem, where $\Lambda$CDM simulations predict dense cuspy centers, in contrary to observations (for review see \cite{core1,core2}). Ultra-light bosonic dark matter (ULDM),  with particle mass $m\sim 10^{-22}\text{eV}$, offers a compelling solution. The macroscopic de Broglie wavelength\cite{ULMD_review},
\[
\lambda_{\text{dB}} \sim 1.92\  \text{kpc} \left( \frac{10^{-22}  \ \text{eV}}{m} \right) \left( \frac{10\  \text{km/s}}{v} \right),
\]
introduces quantum pressure that suppresses small-scale structure and naturally produces flat core-like profiles in dark matter halos\cite{corecusp1,corecusp2,corecusp3}.

Research on ULDM halo structure has diverged into two main approaches: (1) Pioneering work solved the gravitational Schrödinger-Poisson equation numerically to successfully obtain the eigenstate of the  system\cite{nsolve1,nsolve2_decay,nsolve3,nsolve4,nsolve5}.  Although rigorous, these methods are computationally expensive for surveying large galaxy samples; (2) Phenomenological approaches combining various astrophysical probes  suggest a viable mass range of $  10^{-19}\text{eV} >m>10^{-23} \text{eV}$, with Lyman-$\alpha$ data favoring  $m\sim 20\times 10^{-22} \text{eV}$\cite{observe1,observe2,observe3,ob4,ob5,ob6}. However, these often rely on empirical relations rather than a fundamental wavefunction derivation. Additionally, various astrophysical probes place conflicting constraints on $m$.

To bridge the gap between numerically precise but computationally expensive models and phenomenological approaches, this work develops a novel analytic trial wavefunction for ULDM halos. The model is based on the physical premise that in the outskirt of a dwarf galaxy, dark matter self-gravity dominates over particle interaction and baryonic gravitational potential, so the wavefunction should approximate an energy eigenstate.  The eigen-energy is parameterized by a single dimensionless parameter \( t = r_0 / r_{\text{min}} \), where \( r_0 \) is the halo radius and \( r_{\text{min}} \) the BEC core radius. 
We derive the rotation curve  (RC) for the corresponding wavefunction and fit with those for  12 dwarf galaxies (data reference see\cite{data}), with \( m \) as the primary free parameter. Our model achieves excellent fits to all 12 RCs, yielding a consistent particle mass range: $m = (1.8 - 3.2) \times 10^{-23}  \text{eV}$. The remarkable consistency across different galaxies suggests a universal physical scale, strongly supporting the ULDM paradigm.

\section{Schrödinger-Poisson equation of ULDM}

Due to the extremely long wavelength of ULDM, which can be comparable to the scale of galaxies, it is treated as a quantum object rather than classical matter. The Schrödinger-Poisson equation is solved for the ground state of the dark matter. Previous theoretical and numerical works indicate that ULDM forms a BEC in galactic centers, which is responsible for resolving the core-cusp problem\cite{corecusp2,corecusp3,nsolve4}. We assume that particles in the BEC phase do not interfere with those outside of it. At the same time, due to the low density, we also neglect the self-interactions between dark matter particles.  The essence of this problem is given by a many-body  Hamiltonian:
\begin{equation}
H=\sum_i(\frac{\overrightarrow{p}_i^2}{2m}-\frac{GM_b  m}{|\overrightarrow{r_i}|})-\sum_{i<j,}\frac{Gm^2}{|\overrightarrow{r_i}-\overrightarrow{r_j}|},\ \ \ \ \ \ r_i,r_j>r_{\text{min}}
\end{equation}
where $M_b $ is the mass of the BEC, and $\vec{r_i}$ and $\vec{p_i}$ are the position and momentum vectors of the particle. Here we treat the central BEC as a hard core as we assume that there is no interference between the core and outer part of the halo dark matter, except for their gravitational interaction.  After second quantization of the N-body system, the Hamiltonian of the scalar field reads:
\begin{eqnarray}
H=&& \int_{|\overrightarrow{r}|>r_{\text{min}}} d^3r\  \phi(\overrightarrow{r})(-\frac{\hbar^2}{2m}\nabla^2-\mu-\frac{GM_b  m}{r})\phi(\overrightarrow{r}) \nonumber\\
&&-\frac{1}{2}\int_{|\overrightarrow{r}|,|\overrightarrow{r'}|>r_{\text{min}}}  d^3rd^3r'\ \frac{Gm^2}{|\overrightarrow{r}-\overrightarrow{r'}|}\rho(\overrightarrow{r})\rho(\overrightarrow{r'}), \label{H0}
\end{eqnarray}
where $\mu$ is the chemical potential  and $\rho(\overrightarrow{r})$ is the particle number density defined as:
\begin{eqnarray}
\rho(\overrightarrow{r})=\phi(\overrightarrow{r})^2.
\end{eqnarray}
Since we assume a spherically symmetric matter distribution, in this paper we take the scalar field, the spatial part of the energy eigen-function,  to be real. The corresponding  Schrödinger-Poisson equation is
\begin{eqnarray}
(-\frac{\hbar^2}{2m }\nabla^2 -\frac{ GM_b  m}{r})\phi(r)-\int_{|\overrightarrow{r'}|>r_{\text{min}}} d^3r'\ \frac{Gm^2}{|\overrightarrow{r}-\overrightarrow{r'}|}\rho(r')\phi(r)=\mu \phi(r).
\end{eqnarray}
In the following calculations, we propose an analytical trial  wave function and then estimate  its deviation from the true eigenstate. As we will show, although the trial wave function we propose deviates from the true eigenstate significantly  when $r$ is small, the deviation rapidly decays to zero as $r$ increases, yielding the correct RC for the region outside the galactic core, which is what we expect.

\section{Trial wavefunction and numerical results}

 Since the equation is a second-order non-linear integro-differential equation, finding an exact analytic solution is actually impossible. We will adjust the parameters of a plausible trial wave function to minimize its deviation from the true eigenstate to an acceptable range. To begin with, we assume that the wave function is spherically symmetric, which simplifies the original Schrödinger-Poisson equation to:
\begin{eqnarray}
(-\frac{\hbar^2}{2m }\partial_r^2 -\frac{ GM_b  m}{r})u(r)-\frac{G m^2}{r}(\int_{r_{\text{min}}}^r dx \ x^2\rho(x)+r\int_r^{\infty} dx\  x\rho(x))u(r)=\mu u(r), \label{sse}
\end{eqnarray}
where $u(r)=r\phi(r),\ \rho(r)=\phi(r)^2$. For convenience, we set $\hbar=m=G=1$ and we will recover the unit at the end. In regions where $r$ is very large, the galaxy and the dark matter halo can be treated as a point object, and the dark matter wave function should decay exponentially, similar to the eigen wave functions of the hydrogen atom. Let $r_0$ denotes the boundary of the halo, for $r>r_0$,   the wave function is:
\begin{eqnarray}
\phi(r)=n_1 e^{-\eta r}.
\end{eqnarray}
 The corresponding kinetic energy is given by
\begin{eqnarray}
E_k=-\frac{u''}{2u}=-\frac{\eta^2}{2}+\frac{\eta }{r}.
\end{eqnarray}
The total energy of the non-BEC dark matter is then given by $-\eta^2/2$, and its total mass is given by 
\begin{eqnarray}
Nm=\eta-M_b, \label{number}
\end{eqnarray}
where $N$ is the  total number of non-BEC dark matter particles. In the region where $r<r_0$, we assume that 
\begin{eqnarray}
\phi(r)=n_2 r^{-1}e^{-f(r)},
\end{eqnarray}
and the corresponding kinetic energy takes the form:
\begin{eqnarray}
E_k=\frac{1}{2}(f''-f'^2).
\end{eqnarray}
 We find that when
\begin{eqnarray}
f''=\frac{2a}{r}+\frac{2b}{r+c},
\end{eqnarray}
or
\begin{eqnarray}
    f=c_1+c_2 r+2a r(\ln r-1)+2br(\ln(r+c)-1+\frac{c}{r}\ln(r+c)),
\end{eqnarray}
the kinetic energy becomes
\begin{eqnarray}
E_k=\frac{a}{r}+\frac{b}{r+c}-2[c_2+a\ln r+b\ln(r+c)]^2,
\end{eqnarray}
where $c_1,c_2$  are integration constants. Within the  region where $r\ll c$, if $a\ll 1$ and $a c_2\sim 1$, $E_k \sim M_b/r - b\ln r + E$, which gives rise to a constant linear velocity. We acknowledge that this trial wave function will certainly deviate from the true eigen wave function, but it captures the main characteristics of the dark matter distribution. We join the two wave-function segments by matching the wave functions  and their derivatives up to third order at $r=r_0$ to ensure a smooth connection. We have $7$ parameters $\{a,b,c,c_1,c_2,\eta,n_1/n_2 \}$ in total and only four constraint equations.   Finally the wave function takes the form
\begin{eqnarray}
\phi(r)&=& r^{-\beta(r)-1},\nonumber\\
\beta(r)&=& \frac{1-\tanh[g(r-r_0)]}{2}[c_1+c_2 r+2a r(\ln r-1)+2br(\ln(r+c)-1+\frac{c}{r}\ln(r+c))]\nonumber\\
&&+ \frac{\tanh[g(r-r_0)]+1}{2}(\frac{\eta r-n_1/n_2}{\ln r}-1),
\end{eqnarray}
where we combine the two pieces by using the $\tanh$ function (in the numerical calculations, we take $g=1$). $c_1$ is determined by the condition $\beta(r_\text{min})=0$.\footnote{This condition is  intended to make the density $\rho\sim r^{-1}$ in the small $r$ region.} The parameters $\{a,b,\eta,n_1/n_2 \}$ can be solved from the four smooth connection equations. The independent variables are $c,c_2$ for fixed $r_\text{min}$ and $r_0$. The explicit value of $r_\text{min}$ can be modified by rescaling the unit. More importantly, the ratio $t=r_0/r_\text{min}$ determines the size of the dark matter halo and control the (eigen) energy of the system. We will do best-fit for various $t$. We define the deviation of $\phi(r)$ from the eigenstate to be  
\begin{eqnarray}
\delta=\frac{H(r)\phi(r)}{(-\eta^2/2)\phi(r)}-1. \label{deviation}
\end{eqnarray}
We minimize this deviation to find the best fit $c,c_2$ and $M_b $. The linear velocity is determined by the total potential
\begin{eqnarray}
V_t(r)=-\frac{1}{r}[\int_{r_\text{min}}^r dx\ x^2\phi(x)^2+r\int_{r}^{\infty} dx\ x\phi(x)^2]-\frac{M_b }{r}.
\end{eqnarray}

To facilitate numerical computation, we introduce a parameter $\chi$ with unit $L^{-1}$ to render Eq.(\ref{sse}) dimensionless. By defining a dimensionless radius $R=\chi r$ (accordingly $R_\text{min}=\chi r_\text{min},\ R_0=\chi r_0$), we can rewrite the Eq.(\ref{sse}) as:
\begin{eqnarray}
-\frac{u''(R)}{2u(R)}-\frac{GM_b  m^2 }{\hbar^2\chi R}-\frac{8\pi N G m^2}{\hbar^2\chi R}[\int_{R_\text{min}}^R dx\ x^2\phi(x)^2+R\int_{R}^{\infty} dx\ x\phi(x)^2]=\frac{ m \mu  }{\chi^2\hbar^2 }.
\end{eqnarray}
After minimizing $\delta$, we obtain the dimensionless mass of the BEC core $m_b $. The units of radius and velocity can be recovered 
\begin{eqnarray}
\chi=\frac{GM_b  m^2}{m_b  \hbar^2},\ \ \ \ v=v_0\frac{GM_b  m}{m_b  \hbar},  
\end{eqnarray}
where $v_0$ is the dimensionless velocity.  Combining these two relations yields a direct connection between the mass of dark matter particle and the scaling coefficients which is independent of $m_b$:
\begin{eqnarray}
m=\hbar \chi\frac{v_0}{v}.\label{unit}
\end{eqnarray}

\begin{figure}[htbp]
\centering
 \hspace*{-1.0cm}
\includegraphics[width=15cm, height=8cm]{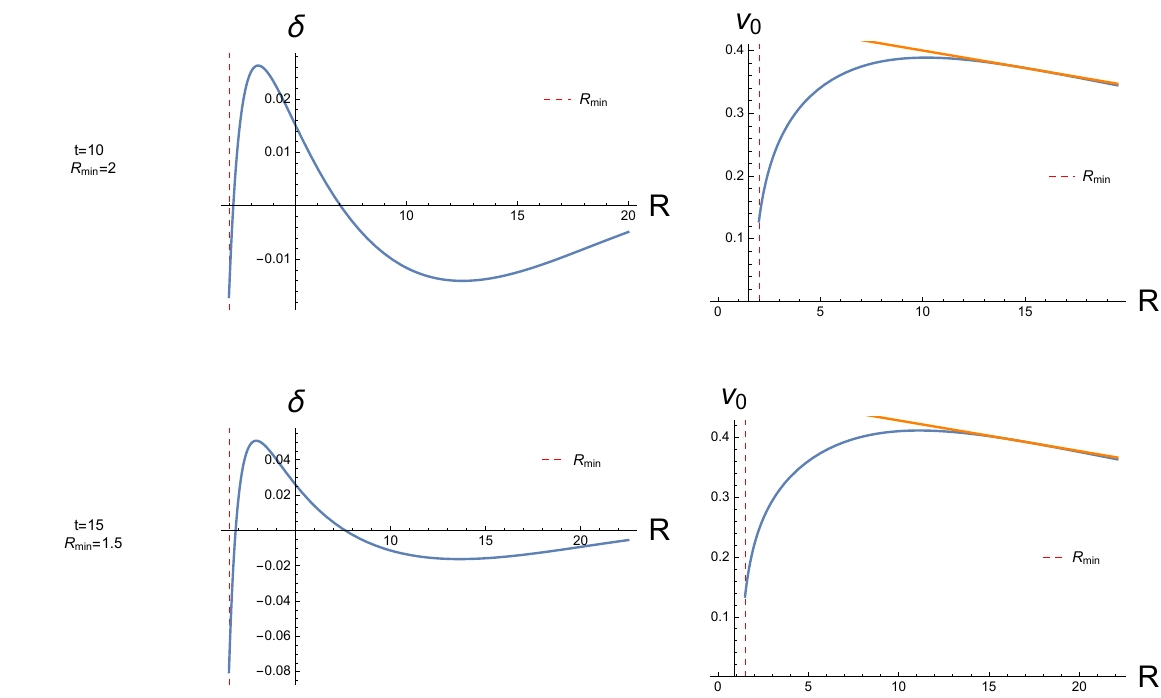}
\caption{Deviation of the trial wave function from the eigen wave function (Eq.(\ref{deviation}), left panel), and the resulting RC (right panel), for $t = 10$ (upper panels) and $t = 15$ (lower panels).}
\label{t10}
\end{figure}

We have plotted $\delta$ (defined in Eq.(\ref{deviation})) and the resulting RCs for different values of $t$, and we list the best-fit parameters in Table \ref{parameters}. In the main text, we only present the results for $t=10$ and $t=15$, as shown in Fig.\ref{t10}; RCs for other values of $t$ will be provided in the Appendix \ref{deviation_and_rotation}. The large $R$ section of the RC is approximately a straight line, and its slope is defined as $k$. We highlight the significant features of these curves:
\begin{itemize}
\item For $t> 20$, $\delta$ is significant for small $R$ but decays rapidly as $R$ increases, becoming almost zero for $R > R_0$.   In the region between $R_\text{min}$ and $R_0$, there is still a substantial range where $\delta$ remains within $5\%$;
\item Due to the significant $\delta$ at small $R$, where the BEC potential contributes more significantly to the energy, there will be considerable bias in the estimation of the BEC mass. This is reflected in the ratio of total mass to BEC mass $\eta/m_b $, which does not increase as expected with increasing $t$. However, the RC is accurate where $\delta$ is small, so the slope of the RC and the total energy of the system will decrease with increasing $t$ as expected;
\item  In our model, the mass and radius of the BEC phase should be fixed, meaning that $R_\text{min}/m_b$ does not change with the system's energy levels. Our calculation results are not consistent with this expectation. This could be due to (1) the estimation of $m_b$ being biased, as mentioned earlier, and (2) the trial wave function lacking a parameter to adjust $R_\text{min}/m_b$. Therefore, the trial wave function we provide might correspond to different systems at different values of $t$. However, because the detailed properties of the central BEC state are unknown and hard to observe directly, this inconsistency is harmless as long as the RC fits the observational data well.

\end{itemize}

\begin{table}
\centering
\caption{The best-fit parameters for different $t$.} \label{parameters}
\begin{tabular}{c c c c c c c c c}
\hline \hline
$t$ & $R_{\min}$ & $c/R_{\text{min}}$ & $c_{2}$ & $m_{b}$ & $\eta$ & $\eta/m_{b}$ & \multicolumn{1}{c}{$k$} \\ 
\hline
10 & 2 & 40 & 0.646 & 0.0327 & 0.149 & 4.56 & -0.00557 \\
15 & 1.5 & 45 & 0.57 & 0.027 & 0.13 & 4.84 & -0.00512 \\
20 & 1.5 & 30 & 0.278 & 0.0247 & 0.105 & 4.25 & -0.00432 \\
30 & 1.5 & 22.5 & 0.141 & 0.0213 & 0.08 & 3.77 & -0.00336 \\
50 & 1.05 & 21 & 0.113 & 0.0217 & 0.07 & 3.24 & -0.00315 \\
100 & 1.05 & 21 & 0.0548 & 0.016 & 0.042 & 2.69 & -0.00167 \\
200 & 5 & 250 & 0.0067 & 0.0014 & 0.05 & 3.23 & -0.0006 \\
\hline
\end{tabular}
\end{table}

\section{Comparison with the observed RCs}

In this section, we fit the RCs of 12 dwarf galaxies using our model as shown in Fig.\ref{dg_fit}. The fitting was performed using the curve$\_$fit package\cite{curve_fit}.
If the dwarf galaxies are mainly dominated by dark matter, our model's RCs agree with observational results over a wide range. However, in the small $r$ regions, the velocities we predict are lower than the observations in these 12 dwarf galaxies. This is because (1) we have not considered the normal matter within the galaxy, and (2) at small $r$, the self-interaction of particles contributes, flattening the density curve to a core.

The resulting best-fit dark matter particle  masses obtained from Eq.(\ref{unit}) range from $1.8\times 10^{-23}\ \text{eV}$ to $3.2\times 10^{-23}\ \text{eV}$ for these dwarf galaxies. These discrepancies are  partly due to the omission of the self-interaction of dark matter and the influence of normal matter. Since the normal matter contributes to the RC, the $v$ in Eq.(\ref{unit}) should be less than the observed value, so the $m$ we obtain is only a lower bound.  By assuming that dwarf galaxies with higher rotational velocities have larger visible matter masses,  the range of $m$ we estimate will be narrowed. However, this approach might be more applicable to larger dwarf galaxies, which are primarily dominated by dark matter. Dwarf galaxies with lower mass might not possess sufficient gravitational force to retain dark matter. It's possible that much of the dark matter  of these galaxies was tidally stripped away.

When fitting the RCs of dwarf galaxies, we utilized only the curves provided by our model for smaller radii (within $r<10r_{\text{min}}$). This is due to the relatively low mass of dwarf galaxies, making it challenging to measure their external RCs. To verify the reliability of our model in regions with larger $r$, we also fitted the RCs of larger galaxies, such as the Milky Way (MW) and M31. Given the complex internal composition of these galaxies, we focused on fitting their external RCs. We present the fitting results in Appendix \ref{galaxy_rotation_curve}, showing good consistency between our model and data.

\section{Conclusion}

This study develops a new analytic model for the density distribution of ULDM in the low-density regions outside the central BEC core. In this regime, we neglect both the self-interactions between ULDM particles and the gravitational potential contributed by baryonic matter—a simplification justified by the dark matter-dominated nature of the dwarf galaxies under study. This leads to a simplified Schrödinger–Poisson system that remains nonlinear due to the self-gravity of the wavefunction. We propose a trial wavefunction designed to approximate the energy eigenstate of this system. Numerical verification confirms that this ansatz exhibits only minimal deviation from the exact solution, particularly in the large $r$ region.

Using the RCs derived from this wavefunction, we perform fits to observational data for 12 dwarf galaxies. Our model achieves excellent agreement with the data and simultaneously constrains the mass of the ULDM particle. The resulting mass estimates across all galaxies are highly consistent, falling within the narrow range of $m=(1.8-3.2)\times 10^{-23}\text{eV}$. These minor variations in the inferred $m$ from individual galaxies may arise from the simplifications of our model, such as neglecting baryonic contributions. Moreover, our model remains valid for the outer regions of ordinary galaxies, predicting that the RCs in these outskirts exhibit the same universal shape. 

This work not only provides new observational support for the ULDM paradigm but also offers a powerful and efficient analytic framework for probing the quantum properties of dark matter on galactic scales, opening new avenues for connecting fundamental physics with astrophysical observations.

\section*{Acknowledgements}

We would like to thank Jian Hu for helpful discussion. This research is supported by grants from the Research Grants Council of the Hong Kong Special Administrative Region, China, under Project No.s AoE/P-404/18 and 14300223.

\appendix

\section{Wave function deviations and RCs result for different $t$} \label{deviation_and_rotation}

\begin{figure}[htbp]
\centering
 \hspace*{-1.0cm}
{\includegraphics[width=15cm, height=21cm]{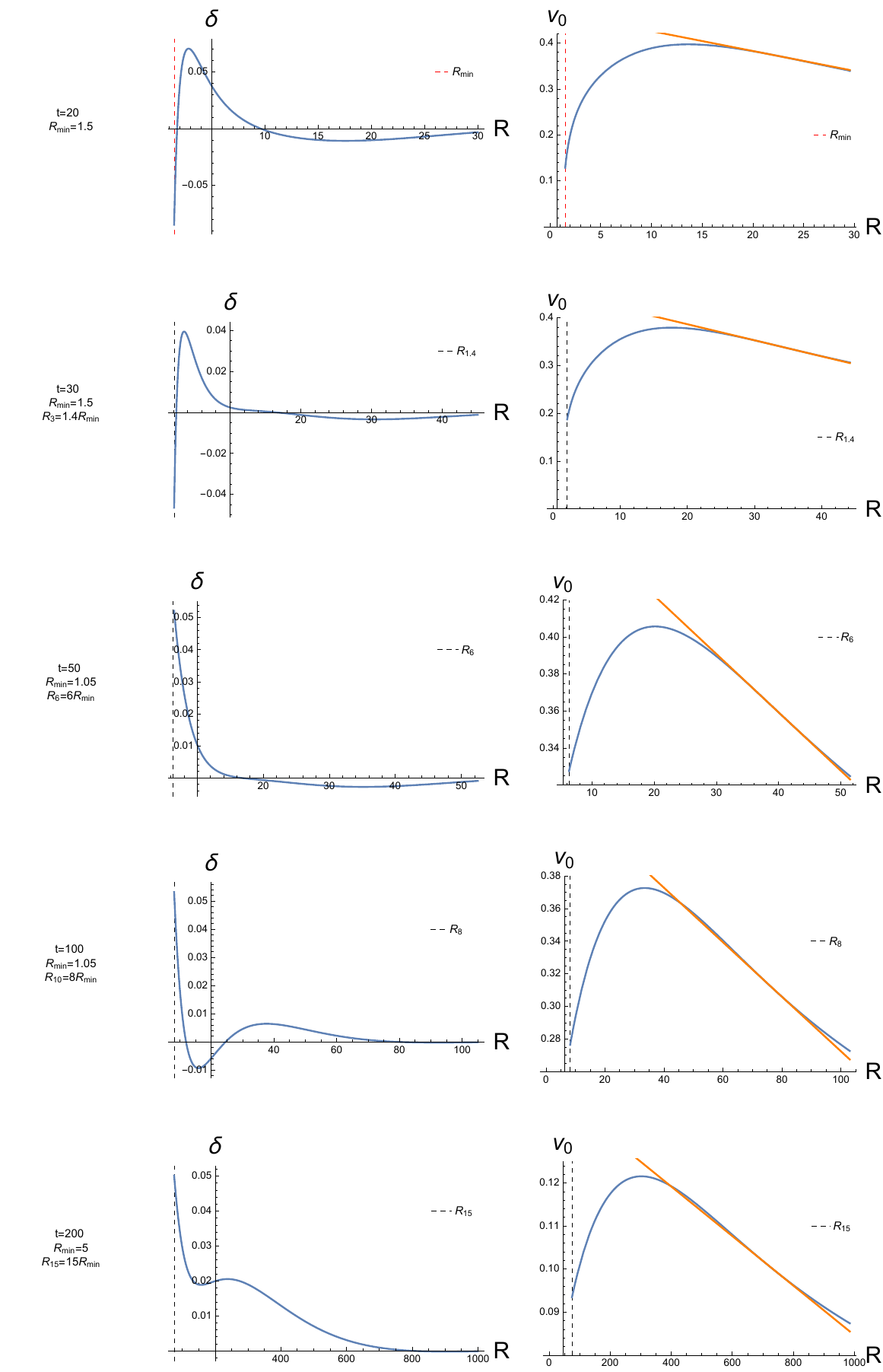}}
\caption{Same as Fig.\ref{t10}, but for $t=20,30,50,100,200$, from top to bottom, respectively. }
\label{t20}
\end{figure}

We provide $\delta$ and RCs for different $t$ in this section as shown in Fig.\ref{t20}. Note that for the RC, since there is significant deviation in the region where $R$ is small, we do not present the RC within this interval. Instead, we only plot the RC where $\delta$ is less than $5\%$. Due to the rapid decay of the wave function beyond $R_0$, the rotation velocity almost takes the form of $R^{-1/2}$. Therefore, we only plot the RC for  $R<R_0$.

\section{Comparison with the RCs of the M31 and MW}\label{galaxy_rotation_curve}

\begin{figure}[htbp]
\centering
\includegraphics[scale=0.6]{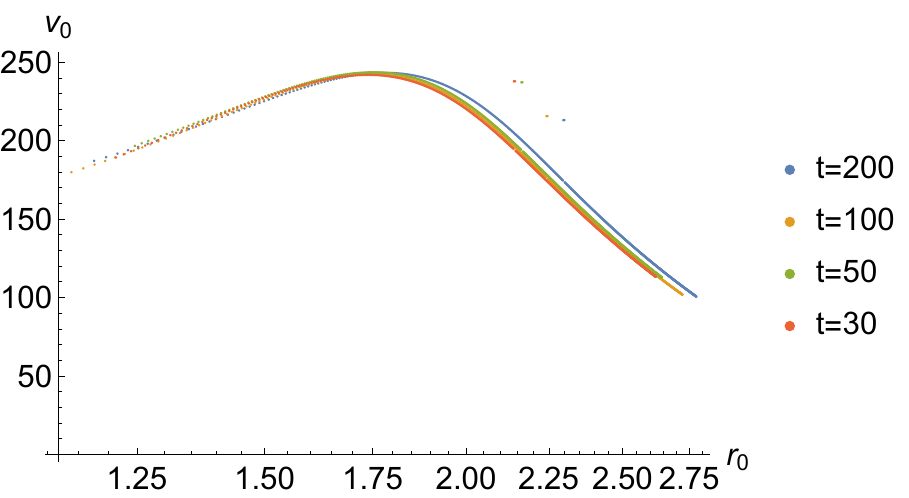}
\caption{RCs for different $t$ and appropriate rescaling of $R$ and $v_0$, showing a  universal shape.}
\label{compare}
\end{figure}

The reason our model can predict the RC outside a galaxy without considering the potential contribution from the normal matter is that, outside the galaxy, the latter also takes the form of $r^{-1}$. If we consider this mass to be included in $M_b$, then the Schrödinger-Poisson equation for the region outside the galaxy is the same as Eq.(\ref{sse}).  However, this does not affect our predictions for the RC in the large  $R$ region because we compare the RCs for different $t$ values after appropriate rescaling of $v_0$ and $r$ as shown in Fig.\ref{compare}. We find that their shapes are almost identical. This suggests that the RCs of all galaxies have this universal shape! The consistency in the shapes of the RCs across different values of $t$ implies that it is independent of the system's radius and mass. Therefore, the RC provided by our model can be used to fit the RCs of the galactic outskirts.

\begin{figure}[htbp]
\centering
\begin{subfigure}{0.45\textwidth}
    \centering
    \includegraphics[width=\linewidth]{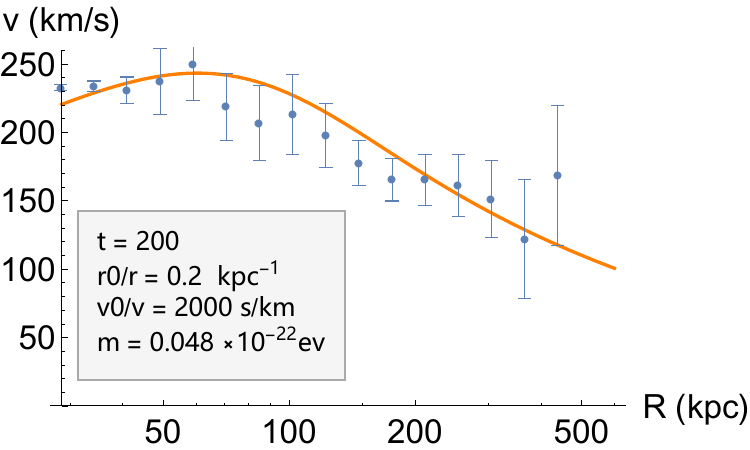}
    \caption{}
\end{subfigure}
\hfill 
\begin{subfigure}{0.45\textwidth}
    \centering
    \includegraphics[width=\linewidth]{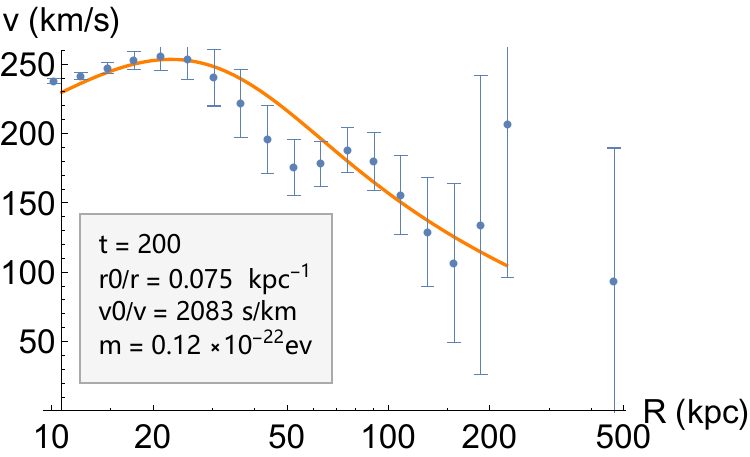}
    \caption{}
\end{subfigure}
\caption{RC of M31 (left panel) and MW (right panel), compared to best-fit model (solid line). The best-fit parameters are shown in the legend.}
\label{M31}
\end{figure}

After appropriate rescaling, we compared our predicted RC with those  of M31 and MW within 500 kpc as shown in Fig.\ref{M31} (data reference see\cite{data2}). Our RC is based on $t=200$, ranging from $30 r_\text{min}$ to $3tr_\text{min}$.  The RCs of these galaxies have the same shape beyond the normal matter boundary (approximately 33 kpc for M31 and 13 kpc for MW), and they are consistent with our predictions.

The difference in the values of $m$ obtained from M31 and MW arises from the fact that the shape of the RC outside the galaxy remains the same across different values of $t$. As $t$ increases, the $m$ derived from Eq.(\ref{unit}) also increases. For example,
\begin{eqnarray}
    m_{30}=0.20m_{200},\ \ m_{50}=0.24m_{200},\ \ m_{100}=0.37m_{200},
\end{eqnarray}
where $m_\alpha$ is obtained based on $t=\alpha$. Without more information about the galactic core, or the value of $t$, we cannot reliably determine the value of $m$. However, the shape of the RC from our model agrees with those of the data in the large $r$ region.

\begin{figure}[htbp]
\centering
 \hspace*{-2.2cm}
\includegraphics[width=19cm, height=20cm]{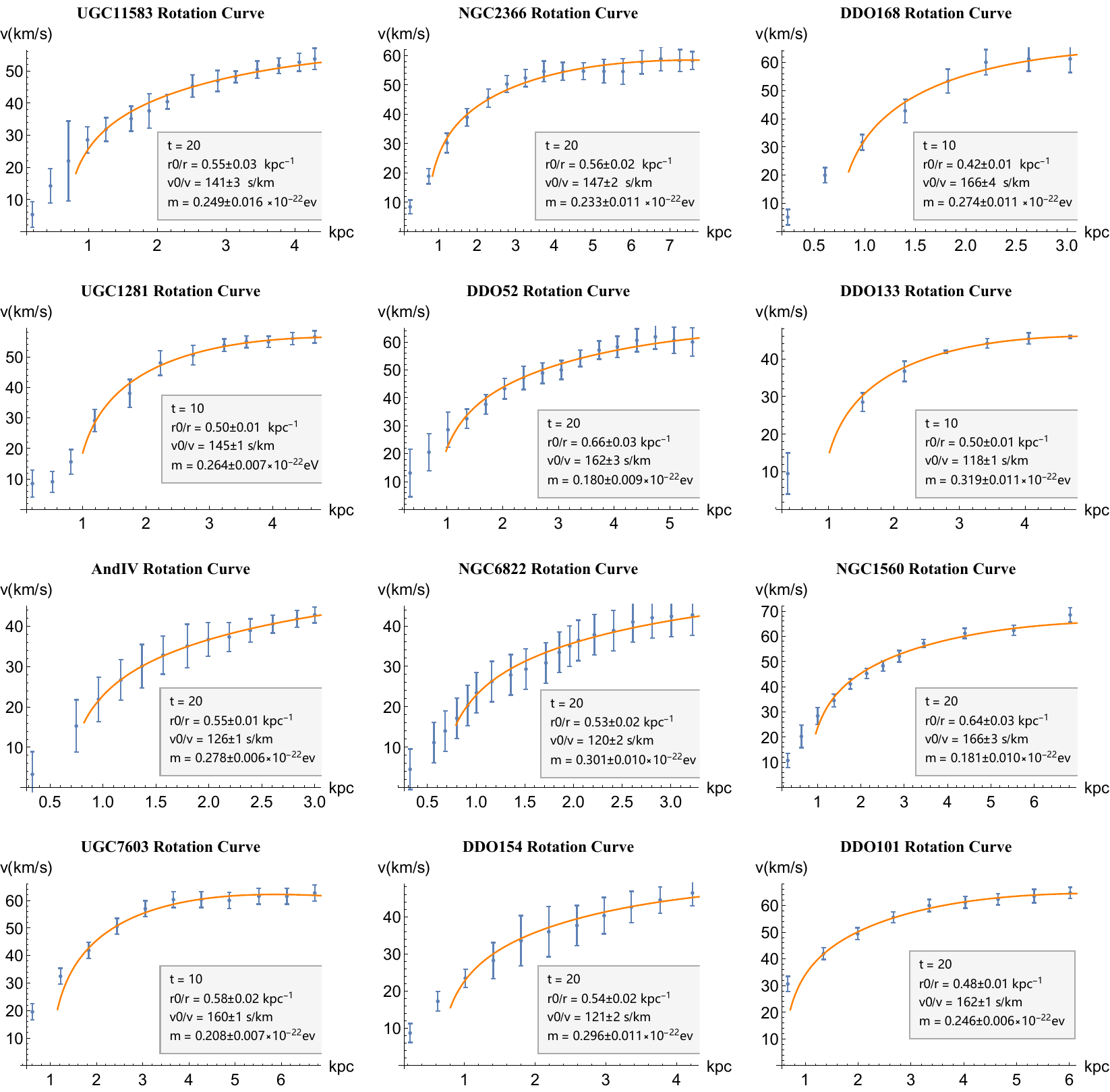}
\caption{Same as Fig.\ref{M31}, but for dwarf galaxies.}
\label{dg_fit}
\end{figure}

\bibliography{ref}

\begin{thebibliography}{21}%
\makeatletter
\providecommand \@ifxundefined [1]{%
 \@ifx{#1\undefined}
}%
\providecommand \@ifnum [1]{%
 \ifnum #1\expandafter \@firstoftwo
 \else \expandafter \@secondoftwo
 \fi
}%
\providecommand \@ifx [1]{%
 \ifx #1\expandafter \@firstoftwo
 \else \expandafter \@secondoftwo
 \fi
}%
\providecommand \natexlab [1]{#1}%
\providecommand \enquote  [1]{``#1''}%
\providecommand \bibnamefont  [1]{#1}%
\providecommand \bibfnamefont [1]{#1}%
\providecommand \citenamefont [1]{#1}%
\providecommand \href@noop [0]{\@secondoftwo}%
\providecommand \href [0]{\begingroup \@sanitize@url \@href}%
\providecommand \@href[1]{\@@startlink{#1}\@@href}%
\providecommand \@@href[1]{\endgroup#1\@@endlink}%
\providecommand \@sanitize@url [0]{\catcode `\\12\catcode `\$12\catcode
  `\&12\catcode `\#12\catcode `\^12\catcode `\_12\catcode `\%12\relax}%
\providecommand \@@startlink[1]{}%
\providecommand \@@endlink[0]{}%
\providecommand \url  [0]{\begingroup\@sanitize@url \@url }%
\providecommand \@url [1]{\endgroup\@href {#1}{\urlprefix }}%
\providecommand \urlprefix  [0]{URL }%
\providecommand \Eprint [0]{\href }%
\providecommand \doibase [0]{http://dx.doi.org/}%
\providecommand \selectlanguage [0]{\@gobble}%
\providecommand \bibinfo  [0]{\@secondoftwo}%
\providecommand \bibfield  [0]{\@secondoftwo}%
\providecommand \translation [1]{[#1]}%
\providecommand \BibitemOpen [0]{}%
\providecommand \bibitemStop [0]{}%
\providecommand \bibitemNoStop [0]{.\EOS\space}%
\providecommand \EOS [0]{\spacefactor3000\relax}%
\providecommand \BibitemShut  [1]{\csname bibitem#1\endcsname}%
\let\auto@bib@innerbib\@empty
\bibitem [{\citenamefont {Pontzen}\ and\ \citenamefont
  {Governato}(2014)}]{core1}%
  \BibitemOpen
  \bibfield  {author} {\bibinfo {author} {\bibfnamefont {A.}~\bibnamefont
  {Pontzen}}\ and\ \bibinfo {author} {\bibfnamefont {F.}~\bibnamefont
  {Governato}},\ }\href {\doibase 10.1038/nature12953} {\bibfield  {journal}
  {\bibinfo  {journal} {Nature}\ }\textbf {\bibinfo {volume} {506}},\ \bibinfo
  {pages} {171–178} (\bibinfo {year} {2014})}\BibitemShut {NoStop}%
\bibitem [{\citenamefont {de~Blok}(2009)}]{core2}%
  \BibitemOpen
  \bibfield  {author} {\bibinfo {author} {\bibfnamefont {W.~J.~G.}\
  \bibnamefont {de~Blok}},\ }\href {\doibase 10.1155/2010/789293} {\bibfield
  {journal} {\bibinfo  {journal} {Advances in Astronomy}\ }\textbf {\bibinfo
  {volume} {2010}} (\bibinfo {year} {2009}),\ 10.1155/2010/789293}\BibitemShut
  {NoStop}%
\bibitem [{\citenamefont {Hui}\ \emph {et~al.}(2017)\citenamefont {Hui},
  \citenamefont {Ostriker}, \citenamefont {Tremaine},\ and\ \citenamefont
  {Witten}}]{ULMD_review}%
  \BibitemOpen
  \bibfield  {author} {\bibinfo {author} {\bibfnamefont {L.}~\bibnamefont
  {Hui}}, \bibinfo {author} {\bibfnamefont {J.~P.}\ \bibnamefont {Ostriker}},
  \bibinfo {author} {\bibfnamefont {S.}~\bibnamefont {Tremaine}}, \ and\
  \bibinfo {author} {\bibfnamefont {E.}~\bibnamefont {Witten}},\ }\href
  {\doibase 10.1103/physrevd.95.043541} {\bibfield  {journal} {\bibinfo
  {journal} {Physical Review D}\ }\textbf {\bibinfo {volume} {95}} (\bibinfo
  {year} {2017}),\ 10.1103/physrevd.95.043541}\BibitemShut {NoStop}%
\bibitem [{\citenamefont {Hu}\ \emph {et~al.}(2000)\citenamefont {Hu},
  \citenamefont {Barkana},\ and\ \citenamefont {Gruzinov}}]{corecusp1}%
  \BibitemOpen
  \bibfield  {author} {\bibinfo {author} {\bibfnamefont {W.}~\bibnamefont
  {Hu}}, \bibinfo {author} {\bibfnamefont {R.}~\bibnamefont {Barkana}}, \ and\
  \bibinfo {author} {\bibfnamefont {A.}~\bibnamefont {Gruzinov}},\ }\href
  {\doibase 10.1103/physrevlett.85.1158} {\bibfield  {journal} {\bibinfo
  {journal} {Physical Review Letters}\ }\textbf {\bibinfo {volume} {85}},\
  \bibinfo {pages} {1158–1161} (\bibinfo {year} {2000})}\BibitemShut
  {NoStop}%
\bibitem [{\citenamefont {Böhmer}\ and\ \citenamefont
  {Harko}(2007)}]{corecusp2}%
  \BibitemOpen
  \bibfield  {author} {\bibinfo {author} {\bibfnamefont {C.~G.}\ \bibnamefont
  {Böhmer}}\ and\ \bibinfo {author} {\bibfnamefont {T.}~\bibnamefont
  {Harko}},\ }\href {\doibase 10.1088/1475-7516/2007/06/025} {\bibfield
  {journal} {\bibinfo  {journal} {Journal of Cosmology and Astroparticle
  Physics}\ }\textbf {\bibinfo {volume} {2007}},\ \bibinfo {pages} {025–025}
  (\bibinfo {year} {2007})}\BibitemShut {NoStop}%
\bibitem [{\citenamefont {Chavanis}(2011)}]{corecusp3}%
  \BibitemOpen
  \bibfield  {author} {\bibinfo {author} {\bibfnamefont {P.-H.}\ \bibnamefont
  {Chavanis}},\ }\href {\doibase 10.1103/PhysRevD.84.043531} {\bibfield
  {journal} {\bibinfo  {journal} {Phys. Rev. D}\ }\textbf {\bibinfo {volume}
  {84}},\ \bibinfo {pages} {043531} (\bibinfo {year} {2011})}\BibitemShut
  {NoStop}%
\bibitem [{\citenamefont {Kaup}(1968)}]{nsolve1}%
  \BibitemOpen
  \bibfield  {author} {\bibinfo {author} {\bibfnamefont {D.~J.}\ \bibnamefont
  {Kaup}},\ }\href {\doibase 10.1103/PhysRev.172.1331} {\bibfield  {journal}
  {\bibinfo  {journal} {Phys. Rev.}\ }\textbf {\bibinfo {volume} {172}},\
  \bibinfo {pages} {1331} (\bibinfo {year} {1968})}\BibitemShut {NoStop}%
\bibitem [{\citenamefont {Guzmán}\ and\ \citenamefont
  {Ureña-López}(2004)}]{nsolve2_decay}%
  \BibitemOpen
  \bibfield  {author} {\bibinfo {author} {\bibfnamefont {F.~S.}\ \bibnamefont
  {Guzmán}}\ and\ \bibinfo {author} {\bibfnamefont {L.~A.}\ \bibnamefont
  {Ureña-López}},\ }\href {\doibase 10.1103/physrevd.69.124033} {\bibfield
  {journal} {\bibinfo  {journal} {Physical Review D}\ }\textbf {\bibinfo
  {volume} {69}} (\bibinfo {year} {2004}),\
  10.1103/physrevd.69.124033}\BibitemShut {NoStop}%
\bibitem [{\citenamefont {RUFFINI}\ and\ \citenamefont
  {BONAZZOLA}(1969)}]{nsolve3}%
  \BibitemOpen
  \bibfield  {author} {\bibinfo {author} {\bibfnamefont {R.}~\bibnamefont
  {RUFFINI}}\ and\ \bibinfo {author} {\bibfnamefont {S.}~\bibnamefont
  {BONAZZOLA}},\ }\href {\doibase 10.1103/PhysRev.187.1767} {\bibfield
  {journal} {\bibinfo  {journal} {Phys. Rev.}\ }\textbf {\bibinfo {volume}
  {187}},\ \bibinfo {pages} {1767} (\bibinfo {year} {1969})}\BibitemShut
  {NoStop}%
\bibitem [{\citenamefont {BERNSTEIN}\ \emph {et~al.}(1998)\citenamefont
  {BERNSTEIN}, \citenamefont {GILADI},\ and\ \citenamefont {JONES}}]{nsolve4}%
  \BibitemOpen
  \bibfield  {author} {\bibinfo {author} {\bibfnamefont {D.~H.}\ \bibnamefont
  {BERNSTEIN}}, \bibinfo {author} {\bibfnamefont {E.}~\bibnamefont {GILADI}}, \
  and\ \bibinfo {author} {\bibfnamefont {K.~R.~W.}\ \bibnamefont {JONES}},\
  }\href {\doibase 10.1142/S0217732398002473} {\bibfield  {journal} {\bibinfo
  {journal} {Modern Physics Letters A}\ }\textbf {\bibinfo {volume} {13}},\
  \bibinfo {pages} {2327} (\bibinfo {year} {1998})},\ \Eprint
  {http://arxiv.org/abs/https://doi.org/10.1142/S0217732398002473}
  {https://doi.org/10.1142/S0217732398002473} \BibitemShut {NoStop}%
\bibitem [{\citenamefont {Harrison}\ \emph {et~al.}(2002)\citenamefont
  {Harrison}, \citenamefont {Moroz},\ and\ \citenamefont {Tod}}]{nsolve5}%
  \BibitemOpen
  \bibfield  {author} {\bibinfo {author} {\bibfnamefont {R.}~\bibnamefont
  {Harrison}}, \bibinfo {author} {\bibfnamefont {I.}~\bibnamefont {Moroz}}, \
  and\ \bibinfo {author} {\bibfnamefont {K.~P.}\ \bibnamefont {Tod}},\ }\href
  {https://arxiv.org/abs/math-ph/0208045} {\enquote {\bibinfo {title} {A
  numerical study of the schrodinger-newton equation 1: Perturbing the
  spherically-symmetric stationary states},}\ } (\bibinfo {year} {2002}),\
  \Eprint {http://arxiv.org/abs/math-ph/0208045} {arXiv:math-ph/0208045
  [math-ph]} \BibitemShut {NoStop}%
\bibitem [{\citenamefont {Iršič}\ \emph {et~al.}(2017)\citenamefont
  {Iršič}, \citenamefont {Viel}, \citenamefont {Haehnelt}, \citenamefont
  {Bolton},\ and\ \citenamefont {Becker}}]{observe1}%
  \BibitemOpen
  \bibfield  {author} {\bibinfo {author} {\bibfnamefont {V.}~\bibnamefont
  {Iršič}}, \bibinfo {author} {\bibfnamefont {M.}~\bibnamefont {Viel}},
  \bibinfo {author} {\bibfnamefont {M.~G.}\ \bibnamefont {Haehnelt}}, \bibinfo
  {author} {\bibfnamefont {J.~S.}\ \bibnamefont {Bolton}}, \ and\ \bibinfo
  {author} {\bibfnamefont {G.~D.}\ \bibnamefont {Becker}},\ }\href {\doibase
  10.1103/physrevlett.119.031302} {\bibfield  {journal} {\bibinfo  {journal}
  {Physical Review Letters}\ }\textbf {\bibinfo {volume} {119}} (\bibinfo
  {year} {2017}),\ 10.1103/physrevlett.119.031302}\BibitemShut {NoStop}%
\bibitem [{\citenamefont {Bozek}\ \emph {et~al.}(2015)\citenamefont {Bozek},
  \citenamefont {Marsh}, \citenamefont {Silk},\ and\ \citenamefont
  {Wyse}}]{observe2}%
  \BibitemOpen
  \bibfield  {author} {\bibinfo {author} {\bibfnamefont {B.}~\bibnamefont
  {Bozek}}, \bibinfo {author} {\bibfnamefont {D.~J.~E.}\ \bibnamefont {Marsh}},
  \bibinfo {author} {\bibfnamefont {J.}~\bibnamefont {Silk}}, \ and\ \bibinfo
  {author} {\bibfnamefont {R.~F.~G.}\ \bibnamefont {Wyse}},\ }\href {\doibase
  10.1093/mnras/stv624} {\bibfield  {journal} {\bibinfo  {journal} {Monthly
  Notices of the Royal Astronomical Society}\ }\textbf {\bibinfo {volume}
  {450}},\ \bibinfo {pages} {209–222} (\bibinfo {year} {2015})}\BibitemShut
  {NoStop}%
\bibitem [{\citenamefont {Fernández-Hernández}\ \emph
  {et~al.}(2019)\citenamefont {Fernández-Hernández}, \citenamefont
  {Montiel},\ and\ \citenamefont {Rodríguez-Meza}}]{observe3}%
  \BibitemOpen
  \bibfield  {author} {\bibinfo {author} {\bibfnamefont {L.~M.}\ \bibnamefont
  {Fernández-Hernández}}, \bibinfo {author} {\bibfnamefont {A.}~\bibnamefont
  {Montiel}}, \ and\ \bibinfo {author} {\bibfnamefont {M.~A.}\ \bibnamefont
  {Rodríguez-Meza}},\ }\href {\doibase 10.1093/mnras/stz1969} {\bibfield
  {journal} {\bibinfo  {journal} {Monthly Notices of the Royal Astronomical
  Society}\ }\textbf {\bibinfo {volume} {488}},\ \bibinfo {pages} {5127–5144}
  (\bibinfo {year} {2019})}\BibitemShut {NoStop}%
\bibitem [{\citenamefont {Hlozek}\ \emph {et~al.}(2015)\citenamefont {Hlozek},
  \citenamefont {Grin}, \citenamefont {Marsh},\ and\ \citenamefont
  {Ferreira}}]{ob4}%
  \BibitemOpen
  \bibfield  {author} {\bibinfo {author} {\bibfnamefont {R.}~\bibnamefont
  {Hlozek}}, \bibinfo {author} {\bibfnamefont {D.}~\bibnamefont {Grin}},
  \bibinfo {author} {\bibfnamefont {D.~J.~E.}\ \bibnamefont {Marsh}}, \ and\
  \bibinfo {author} {\bibfnamefont {P.~G.}\ \bibnamefont {Ferreira}},\ }\href
  {\doibase 10.1103/PhysRevD.91.103512} {\bibfield  {journal} {\bibinfo
  {journal} {Phys. Rev. D}\ }\textbf {\bibinfo {volume} {91}},\ \bibinfo
  {pages} {103512} (\bibinfo {year} {2015})}\BibitemShut {NoStop}%
\bibitem [{\citenamefont {Bar}\ \emph {et~al.}(2018)\citenamefont {Bar},
  \citenamefont {Blas}, \citenamefont {Blum},\ and\ \citenamefont
  {Sibiryakov}}]{ob5}%
  \BibitemOpen
  \bibfield  {author} {\bibinfo {author} {\bibfnamefont {N.}~\bibnamefont
  {Bar}}, \bibinfo {author} {\bibfnamefont {D.}~\bibnamefont {Blas}}, \bibinfo
  {author} {\bibfnamefont {K.}~\bibnamefont {Blum}}, \ and\ \bibinfo {author}
  {\bibfnamefont {S.}~\bibnamefont {Sibiryakov}},\ }\href {\doibase
  10.1103/PhysRevD.98.083027} {\bibfield  {journal} {\bibinfo  {journal} {Phys.
  Rev. D}\ }\textbf {\bibinfo {volume} {98}},\ \bibinfo {pages} {083027}
  (\bibinfo {year} {2018})}\BibitemShut {NoStop}%
\bibitem [{\citenamefont {Zimmermann}\ \emph {et~al.}(2025)\citenamefont
  {Zimmermann}, \citenamefont {Alvey}, \citenamefont {Marsh}, \citenamefont
  {Fairbairn},\ and\ \citenamefont {Read}}]{ob6}%
  \BibitemOpen
  \bibfield  {author} {\bibinfo {author} {\bibfnamefont {T.}~\bibnamefont
  {Zimmermann}}, \bibinfo {author} {\bibfnamefont {J.}~\bibnamefont {Alvey}},
  \bibinfo {author} {\bibfnamefont {D.~J.~E.}\ \bibnamefont {Marsh}}, \bibinfo
  {author} {\bibfnamefont {M.}~\bibnamefont {Fairbairn}}, \ and\ \bibinfo
  {author} {\bibfnamefont {J.~I.}\ \bibnamefont {Read}},\ }\href {\doibase
  10.1103/PhysRevLett.134.151001} {\bibfield  {journal} {\bibinfo  {journal}
  {Phys. Rev. Lett.}\ }\textbf {\bibinfo {volume} {134}},\ \bibinfo {pages}
  {151001} (\bibinfo {year} {2025})}\BibitemShut {NoStop}%
\bibitem [{\citenamefont {Karukes}\ and\ \citenamefont {Salucci}(2016)}]{data}%
  \BibitemOpen
  \bibfield  {author} {\bibinfo {author} {\bibfnamefont {E.~V.}\ \bibnamefont
  {Karukes}}\ and\ \bibinfo {author} {\bibfnamefont {P.}~\bibnamefont
  {Salucci}},\ }\href {\doibase 10.1093/mnras/stw3055} {\bibfield  {journal}
  {\bibinfo  {journal} {Monthly Notices of the Royal Astronomical Society}\
  }\textbf {\bibinfo {volume} {465}},\ \bibinfo {pages} {4703–4722} (\bibinfo
  {year} {2016})}\BibitemShut {NoStop}%
\bibitem [{Note1()}]{Note1}%
  \BibitemOpen
  \bibinfo {note} {This condition is intended to make the density $\rho \sim
  r^{-1}$ in the small $r$ region.}\BibitemShut {Stop}%
\bibitem [{\citenamefont {{Vugrin}}\ \emph {et~al.}(2007)\citenamefont
  {{Vugrin}}, \citenamefont {{Swiler}}, \citenamefont {{Roberts}},
  \citenamefont {{Stucky-Mack}},\ and\ \citenamefont {{Sullivan}}}]{curve_fit}%
  \BibitemOpen
  \bibfield  {author} {\bibinfo {author} {\bibfnamefont {K.~W.}\ \bibnamefont
  {{Vugrin}}}, \bibinfo {author} {\bibfnamefont {L.~P.}\ \bibnamefont
  {{Swiler}}}, \bibinfo {author} {\bibfnamefont {R.~M.}\ \bibnamefont
  {{Roberts}}}, \bibinfo {author} {\bibfnamefont {N.~J.}\ \bibnamefont
  {{Stucky-Mack}}}, \ and\ \bibinfo {author} {\bibfnamefont {S.~P.}\
  \bibnamefont {{Sullivan}}},\ }\href {\doibase 10.1029/2005WR004804}
  {\bibfield  {journal} {\bibinfo  {journal} {Water Resources Research}\
  }\textbf {\bibinfo {volume} {43}},\ \bibinfo {eid} {W03423} (\bibinfo {year}
  {2007})}\BibitemShut {NoStop}%
\bibitem [{\citenamefont {Sofue}(2015)}]{data2}%
  \BibitemOpen
  \bibfield  {author} {\bibinfo {author} {\bibfnamefont {Y.}~\bibnamefont
  {Sofue}},\ }\href {\doibase 10.1093/pasj/psv042} {\bibfield  {journal}
  {\bibinfo  {journal} {Publications of the Astronomical Society of Japan}\
  }\textbf {\bibinfo {volume} {67}} (\bibinfo {year} {2015}),\
  10.1093/pasj/psv042}\BibitemShut {NoStop}%
\end{thebibliography}%

\end{document}